\documentclass{iaus}
\usepackage{graphicx}

\def\degrees{$^{\circ}$}
\def\ergss{ergs~s$^{-1}$}
\def\sqig{$\sim$}

\title[X-ray Binaries in the Magellanic Clouds] 
{Properties of X-ray Binaries in the Magellanic Clouds from RXTE and
Chandra Observations}

\author[R.H.D. Corbet et al.]   
{R.H.D. Corbet$^{1,2}$, M.J. Coe$^3$, K.E. McGowan$^3$, M.P.E. Schurch$^3$,
L.J. Townsend$^3$, J.L. Galache$^4$, \and F.E. Marshall$^2$}

\affiliation{$^1$ University of Maryland, Baltimore County/CRESST, 
1000 Hilltop Circle, Baltimore, MD 21250, USA\\[\affilskip]
$^2$NASA Goddard Space Flight Center, Greenbelt, MD 20771, USA\\[\affilskip]
$^3$School of Physics and Astronomy, University of Southampton,
SO17 1BJ, UK\\[\affilskip]
$^4$Harvard-Smithsonian Center for Astrophysics, 60 Garden Street, Cambridge, MA 02138, USA\\[\affilskip]
}

\pubyear{2008}
\volume{256}  
\pagerange{1--6}
\jname{The Magellanic System: Stars, Gas, and Galaxies}
\editors{Jacco Th. van Loon \& Joana M. Oliveira, eds.}
\begin{document}

\maketitle

\begin{abstract}
The X-ray binary population of the SMC is very different from that of
the Milky Way consisting, with one exception, entirely of transient
pulsating Be/neutron star binaries.  We have now been monitoring these
SMC X-ray pulsars for over 10 years using the Rossi X-ray Timing
Explorer with observations typically every week. The RXTE observations
have been complemented with surveys made using the Chandra
observatory.  The RXTE observations are non-imaging but enable
detailed studies of pulsing sources.  In contrast, Chandra
observations can provide precise source locations and detections of
sources at lower flux levels, but do not provide the same timing
information or the extended duration light curves that RXTE
observations do. We summarize the results of these monitoring programs
which provide insights into both the differences between the SMC and
the Milky Way, and the details of the accretion processes in X-ray
pulsars.
\end{abstract}

\firstsection 
\section{Introduction}

Mass transfer in high-mass X-ray binaries (HMXBs)
may occur in 3 different ways from the OB star component. (i) The mass-donor primary
star may fill its Roche lobe. These systems are very luminous ($\sim$10$^{38}$ \ergss) but
are very rare.  (ii) If the system contains a supergiant primary with an
extensive stellar wind
then accretion from the wind may take place. These systems have modest
luminosity ($\sim$10$^{36}$ - 10$^{37}$ \ergss) but are rather
more common. (iii) For systems containing a Be
star accretion takes place from the circumstellar envelope. These have a wide
range of luminosities
(10$^{34}$ - 10$^{39}$ \ergss) and are very common, but are transient.

In most HMXBs the accreting object is a highly magnetized neutron
star. Accretion is funneled onto the magnetic poles of the neutron
star and we see pulsations at the neutron star spin period.
If the pulse periods of HMXBs are plotted against their
corresponding orbital periods then it is seen that sources
divide into three groups in this diagram which correspond
to the three modes of mass transfer (Corbet 1986).
In particular there is strong correlation between pulse
period and orbital period for the Be star systems.
The positions of sources in this diagram is thought to depend
on the accretion torques experienced by the
neutron stars and hence on the circumstellar environments
around the primary stars.
These classes of HMXB are well-studied in the Galaxy
and we wish to know how the HXMB populations compare
in other galaxies. Because of their proximity, the
SMC and LMC make them the easiest external galaxies
to investigate.

\begin{figure}[]
\begin{center}
 \includegraphics[width=3.2in,angle=-90]{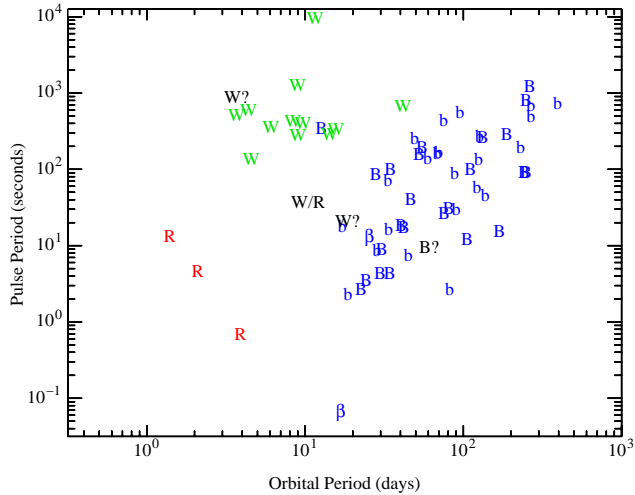}
 \caption{Pulse period vs. orbital period for HMXBs. ``R''
= Roche lobe overflow, ``W'' = wind accretion, ``B'' =
Galactic Be star source, ``b'' = SMC Be star source,
``$\beta$'' = LMC Be star source.}
   \label{fig1}
\end{center}
\end{figure}

Initial estimates of the estimated HMXB population
of the SMC were based on the mass of the SMC.
The SMC is a few percent of the mass of the Galaxy
and about 65 Galactic X-ray pulsars are known.
Therefore, 1 or 2 X-ray pulsars would be expected in the SMC.
The larger fraction of Be stars in the SMC increased the estimate to \sqig3.
The first X-ray pulsar discovered in the SMC was SMC X-1 in 1970s.
Its luminosity can reach \sqig10$^{39}$ \ergss and it has a
0.71s pulse period and a 3.89 day orbital period.
The mass-donating companion is a Roche-lobe filling B0I star.
In 1978 two transients, SMC X-2 and SMC X-3, were found (Clark \etal\
1978).
The three pulsars then known agreed with the
simple prediction, although all three were surprisingly bright.

\section{RXTE Observations of the SMC}
 
RXTE was launched in 1995 and its primary instrument is the
Proportional Counter Array (PCA).  The RXTE PCA has a 2\degrees\ FWZI,
1\degrees\ FWHM field of view.  The PCA is non-imaging, but it has a large
collecting area of up to 7,000 cm$^2$.  The RXTE observing program is
extremely flexible and almost all observations are time constrained.
These include monitoring, phase constrained, and target of opportunity
observations as well as observations coordinated with other
observatories both in space and ground-based.

Serendipitous RXTE PCA slew observations in 1997 showed a possible
outburst from SMC X-3 (Marshall \etal\ 1997). A follow-up pointed RXTE
observation showed a complicated power spectrum with several
harmonic, almost-harmonic, and non-harmonic peaks. Imaging ASCA
observations were then made of this region and they
showed the presence of two separate pulsars. However, neither
of these pulsars coincided with the position of 
SMC X-3.  A revised look at the RXTE power spectrum revealed three
pulsars simultaneously active with periods of 46.6, 91.1, and 74.8 s
(Corbet \etal\ 1998).

Since 1997 we have monitored one or more positions weekly using the
RXTE PCA.  
The flexible
observing program of RXTE has enabled us to carry out a regular monitoring
program that would not have been possible with other satellites.
The typical observation duration has been about 10,000 seconds.  We
use power spectra of the light curves to extract pulsed flux from any
X-ray pulsars in the FOV.  The sensitivity to pulsed flux is
$\sim$10$^{36}$ \ergss\ at the distance of the SMC. From
this program we have detected
many transient sources and all identified optical counterparts have
been found to be Be stars. The SMC HMXB pulsar population has
now been found by ourselves and other investigators to be much
larger than originally thought.
Our naming convention for SMC pulsars is SXPx, where ``x'' is
the pulse period, for {\it SMC X-ray Pulsar}. This convention
is particularly useful for X-ray pulsars discovered with RXTE
for which a precise position is not yet available.
For detailed light curves and
analyses see \cite{Laycock05} and \cite{Galache08}.
In addition, we have recently been able to measure orbital
parameters from Doppler modulation of the pulse period
of SXP18.3 (Schurch \etal\ 2008).

\begin{figure}[]
\begin{center}
 \includegraphics[width=3.0in]{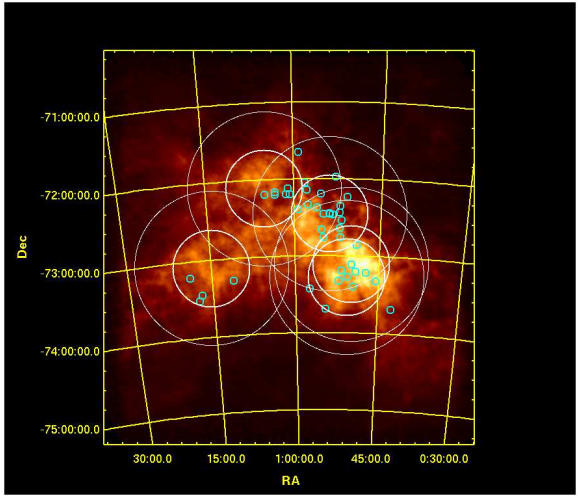}
 \caption{HI Image of the SMC. Large circles = PCA FOV (FWHM and FWZI)
at different monitoring positions.
Small circles show locations of X-ray pulsars.}
   \label{fig2}
\end{center}
\end{figure}

\begin{figure}[]
\begin{center}
 \includegraphics[width=3.4in]{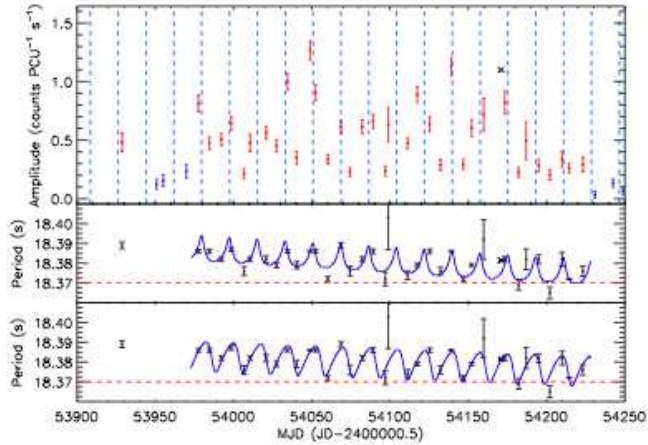}
 \caption{The extended outburst from SXP 18.3.
The top panel shows the amplitude of the pulsed
flux. The two lower panels show two possible
timing solutions. The middle panel shows the
preferred solution with the orbital period
fixed at the photometric period.
(Schurch \etal\ 2008).}
   \label{fig3}
\end{center}
\end{figure}

 
The Be pulsar spin period/orbital period correlation is believed to be
related to the structure of the extended envelopes of Be stars.  SMC
and Milky Way Be stars have differences, for example, the SMC
metallicity is far lower and the Be phenomenon is more common in the SMC.
Is this reflected in the P$_s$/P$_{orb}$ relation? That is, are there
significant differences between Be star envelopes in the SMC and the
Galaxy?  For a linear fit (to the log-log diagram) the intercept is
related to Be star mass loss rates and the gradient is related to the
radial structure of Be star envelopes.

Currently 23 SMC Be X-ray pulsars now have measured orbital periods. 
The periods have been measured by several techniques.
These include: X-ray flux monitoring with RXTE,
pulse timing with RXTE (one system) and
optical observations from MACHO and OGLE.
In comparison, 24 Galactic Be X-ray pulsars now
have measured orbital periods.
We find that for the SMC and Galactic systems the intercepts are the
same, the gradients are the same, and the scatter about the fits
are the same.
Thus, the metallicity difference between the two galaxies gives no
measurable effect on the spin period/orbital period relationship and
the Be star envelopes in SMC and Galaxy are apparently similar.

\begin{figure}[]
\begin{center}
 \includegraphics[width=2.5in,angle=0]{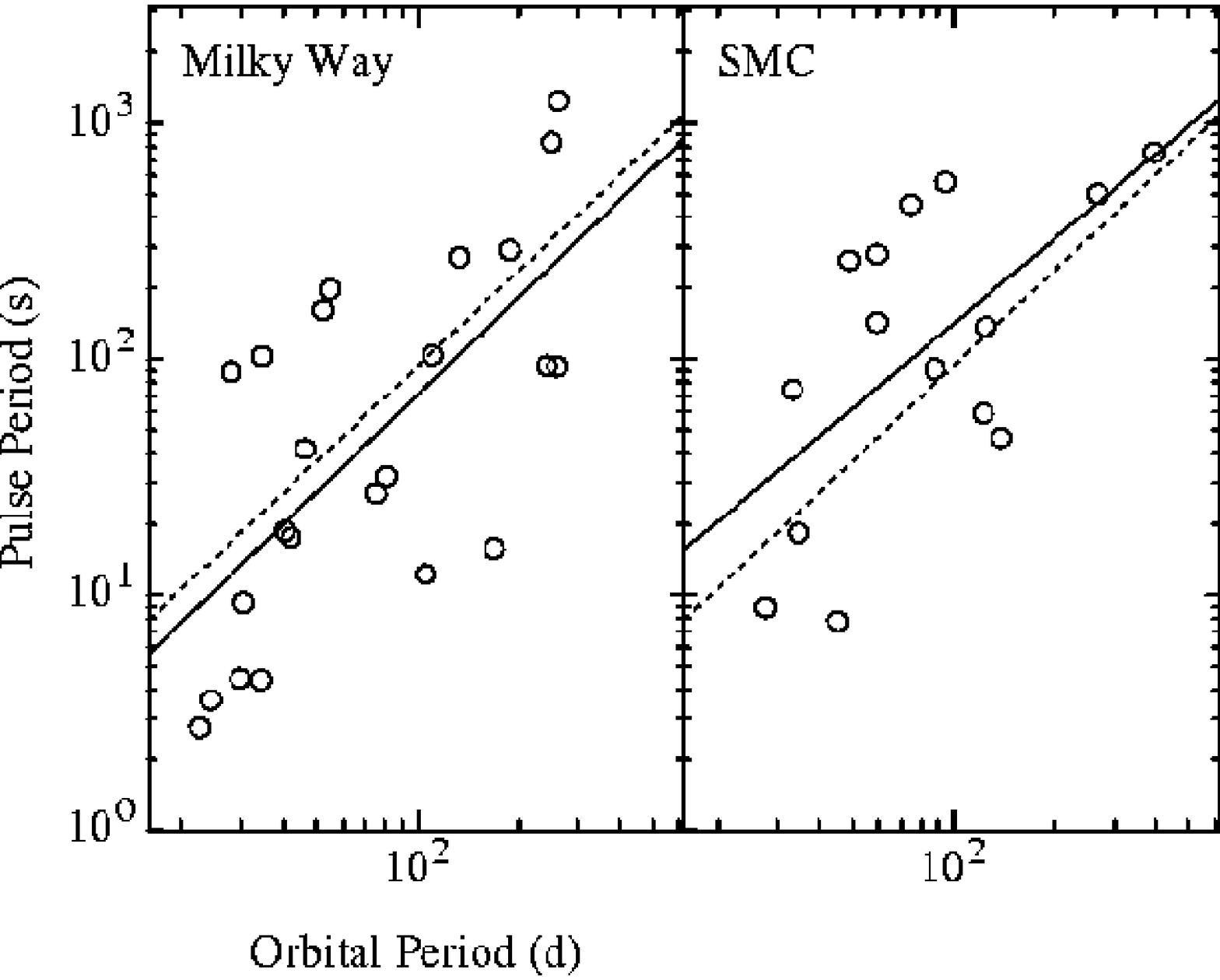}
 \includegraphics[width=2.5in]{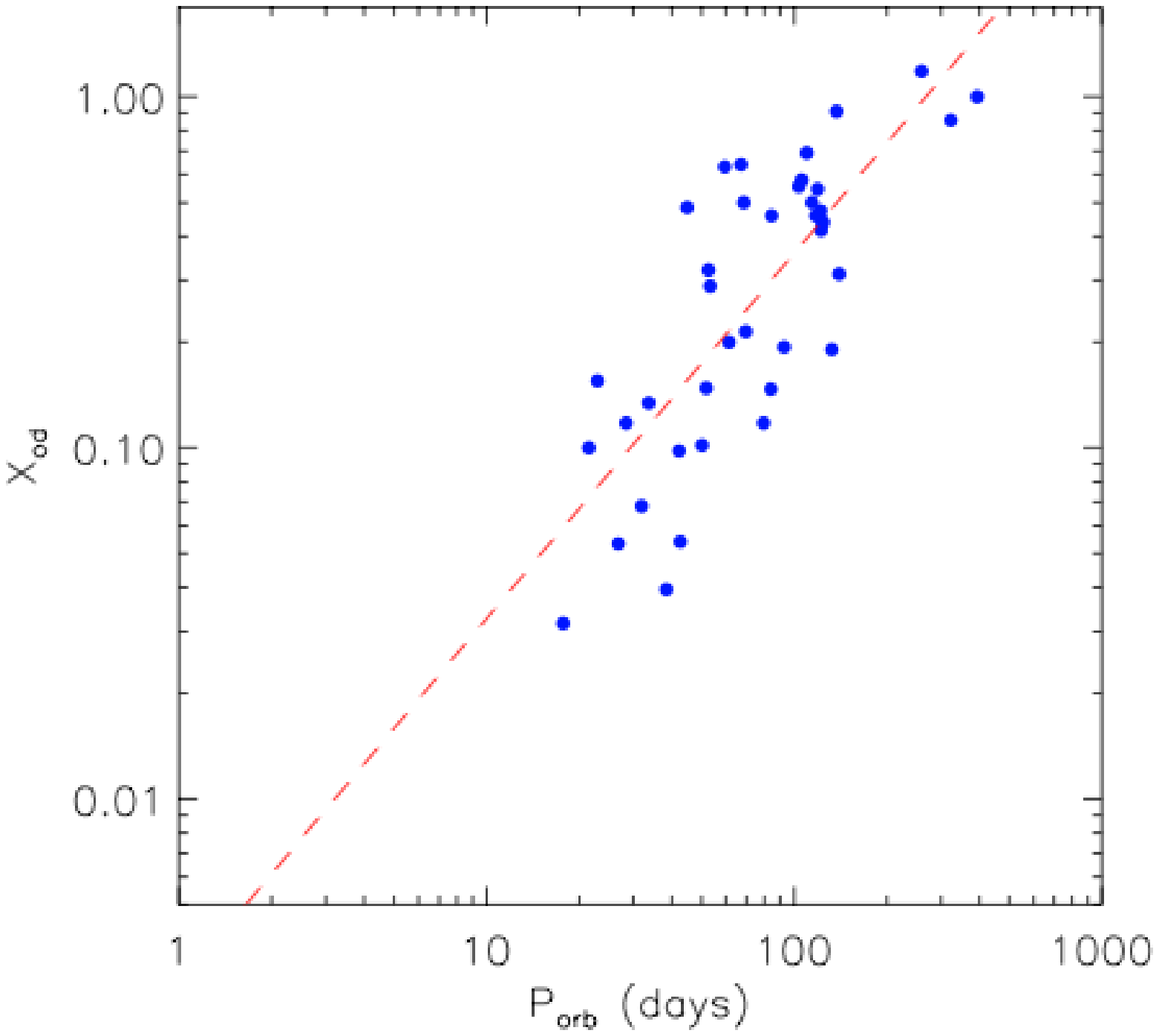}
 \caption{{\it Left:} A comparison of the P$_s$/P$_{orb}$
relationships for the SMC and the Galaxy.
{\it Right:} The relationship between outburst density
and orbital period proposed by \cite{Galache06}.}
   \label{fig4}
\end{center}
\end{figure}

\cite{Galache06} proposes that the frequency of outbursts per orbit
(X-ray ``outburst density'' or X$_{od}$) depends on the orbital period.
Long period systems are more likely to show an outburst at periastron.
The reason for this correlation is not yet clear.

\section{Chandra SMC Wing Survey}
A possible connection between hydrogen column density
(N$_H$) and HMXB location was proposed by \cite{Coe05}.
To investigate this we undertook
a survey of the SMC wing using Chandra.
We observed
20 fields with $\sim$10ks observation time per field.  
523 sources were detected,(\cite{McGowan08} but
only $\sim$5 of these were HMXBs (\cite{McGowan07}) 
and the majority of sources
are probably background AGNs.  
There thus appear to be fewer X-ray pulsars in the wing than the bar.
This is despite that fact that the most luminous SMC HMXB, SMC X-1
is located in the wing.

\begin{figure}[]
\begin{center}
 \includegraphics[width=2.5in]{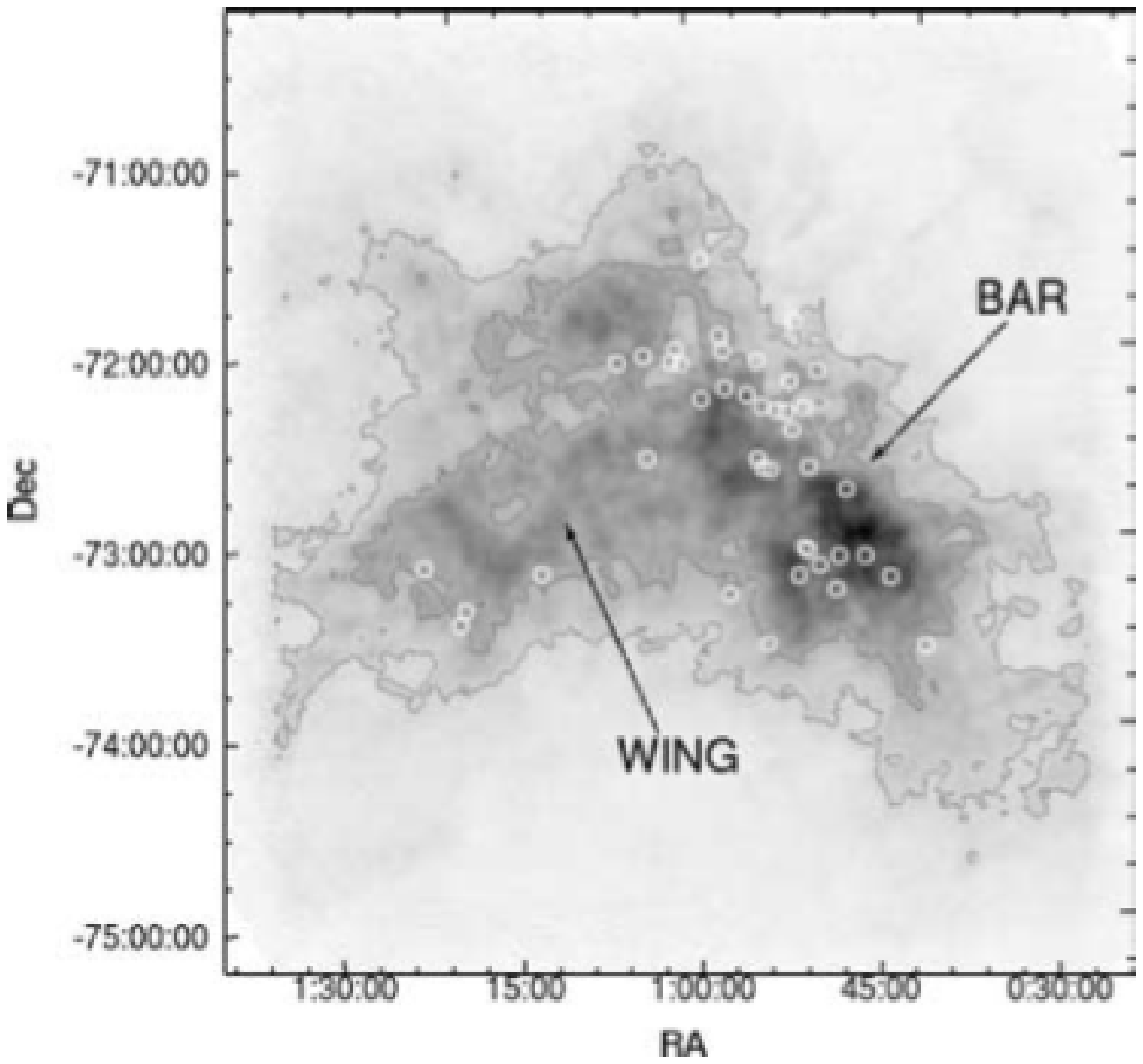}
 \includegraphics[width=2.5in]{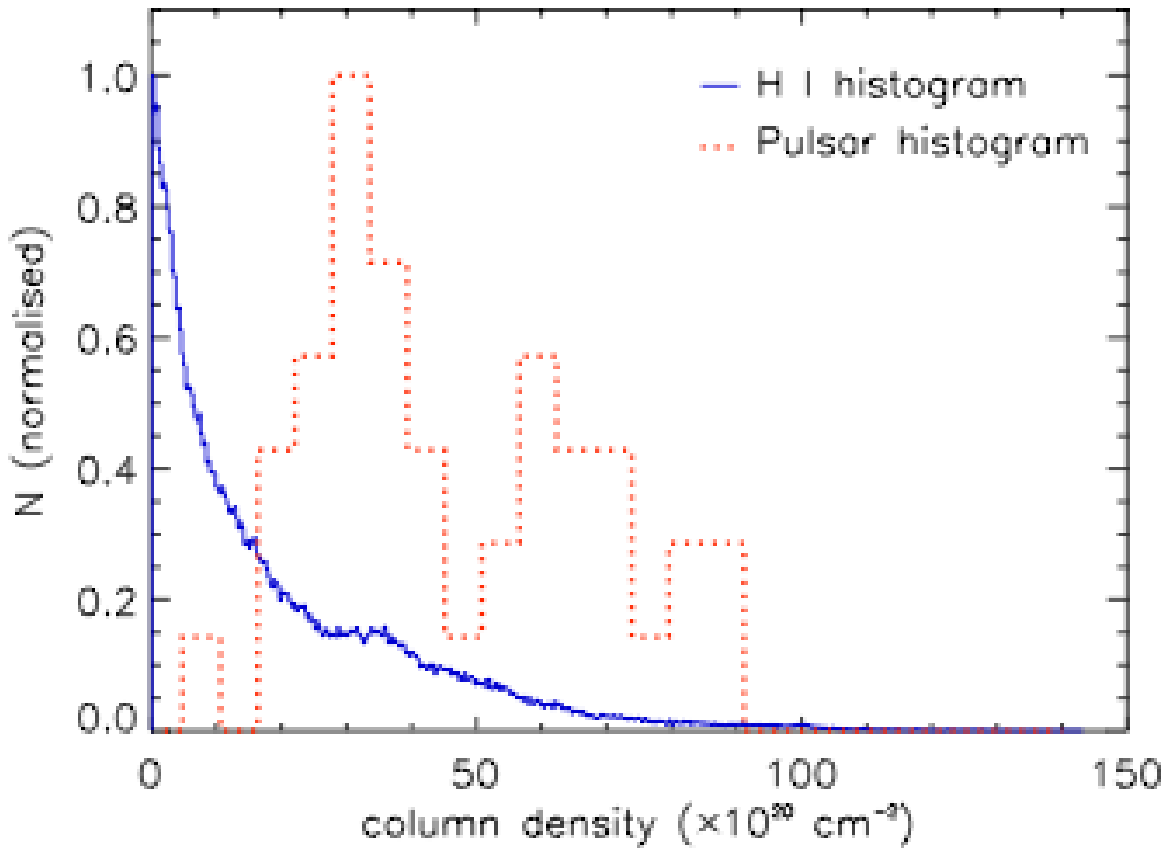}
 \caption{{\it Left:} The location of SMC pulsars
superimposed on an H I contour map. {\it Right:}
Histogram of SMC H I distributions and corresponding
histogram of H I columns at the location of the
X-ray pulsars (Coe \etal\ 2005).}
   \label{fig5}
\end{center}
\end{figure}

\section{RXTE Monitoring of the LMC}
The SMC appears to be very abundant in Be X-ray pulsars. This was
only known after regular observations of the SMC started. The known
LMC X-ray pulsar population is more modest. There
is one Roche lobe overflow
source, and a few Be systems. 
To investigate the LMC population in more
detail we undertook an RXTE monitoring program similar
to the one used for the SMC.  However, the angular
size of the LMC is larger so
we restricted the program to monitoring
one position that was already know to contain several X-ray
sources.
We analyzed data from our one year
monitoring program, together with archival data from other programs
(Townsend et al., in preparation).  In the monitoring region 4 of 
the 5 known
X-ray pulsars were detected.  However, no new X-ray pulsars were discovered.
This implies that the X-ray pulsar content of the LMC is more like
that of the Galaxy than the SMC.

\begin{figure}[]
\begin{center}
 \includegraphics[width=2.5in]{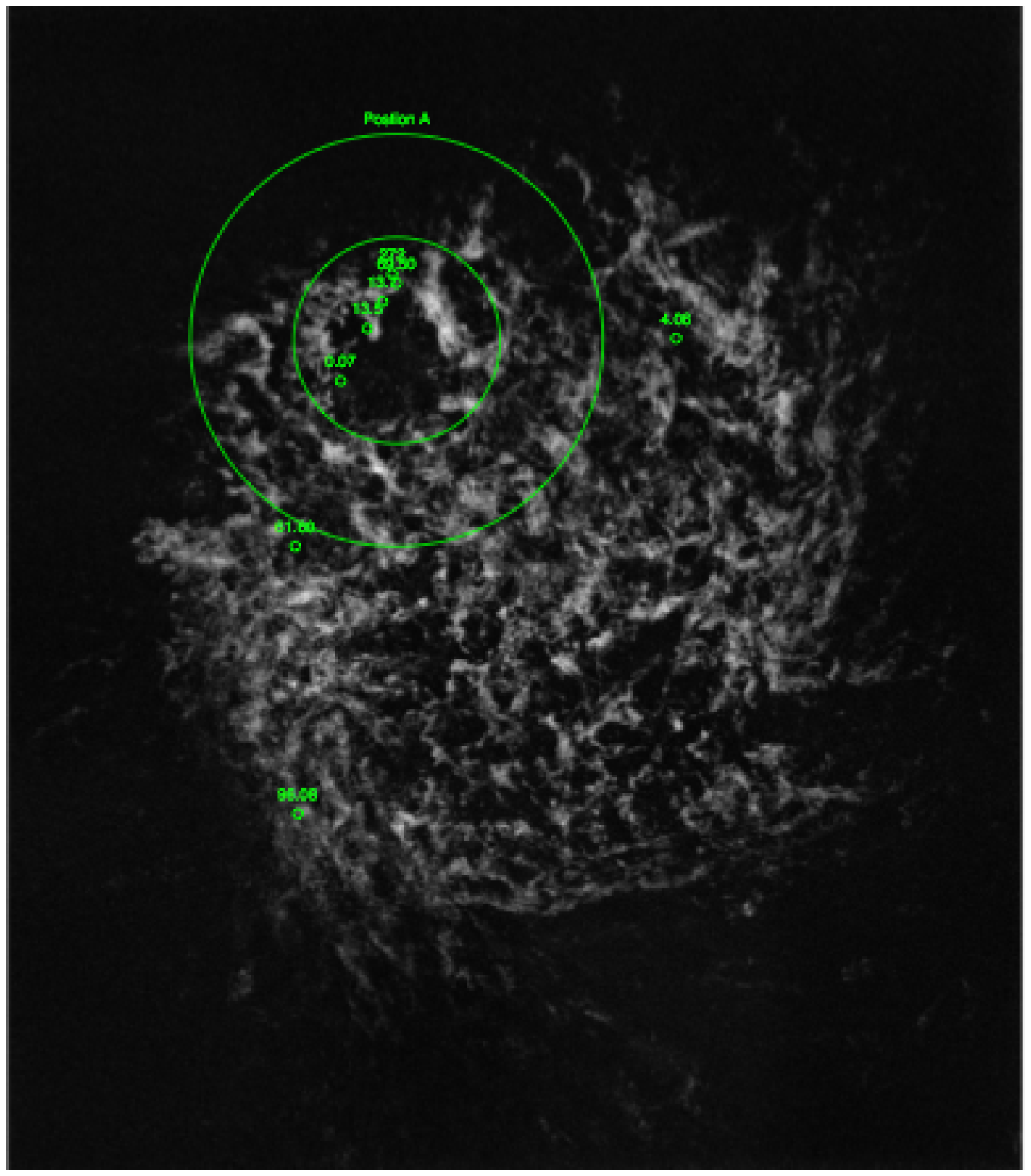}
 \caption{The PCA monitoring position for the LMC. Known pulsar
positions are marked.}
   \label{fig6}
\end{center}
\end{figure}

\section{Conclusion}
The current census of SMC X-ray pulsars is:
1 supergiant Roche lobe filler (SMC X-1);  $\sim$50 transients (likely
all Be star systems);  1 possible Crab-like pulsar (P = 0.087s) from
ASCA (Yokogawa \& Koyama 2000);  1 Anomalous X-ray Pulsar (AXP)
candidate (P = 8.02s) from Chandra and XMM; no supergiant wind
accretion systems and no low-mass X-ray binaries.. 
Supergiant wind systems should easily be detectable at
our $\sim$10$^{36}$ \ergss\
pulsed flux sensitivity.
An obvious question is: why are there so many SMC X-ray pulsars? 
The current star formation rate in the SMC
is reported not to be extremely high.  
The lifetime of HMXBs is short which implies an enhanced
star formation rate in the recent past.  However,
supergiant wind systems,
which have even shorter lifetimes than Be star systems, 
have not been found.
Models of historic star formation rates in the SMC and LMC must be
compatible with the observed X-ray binary populations, and they
most also account for the
differences between the SMC and LMC.

There are also similarities between the SMC and Galactic
pulsar populations.
The SMC and Galactic
Be star systems have
identical
(within errors) P$_s$/P$_{orb}$ relationships.  
The LMC X-ray pulsar population also
appears to be more similar to that of the Galaxy.  
The large and growing SMC X-ray pulsar database
has considerable
potential for understanding the astrophysics of accretion
processes. 
It facilitates comparative studies, such
as pulse profile morphology, as a
function of luminosity. Or, luminosity effects can
be removed and we can examine the effects of
other parameters such as magnetic field strength.  
The SMC is nearby and optical counterparts
can be observed with modest size telescopes. 
In particular, MACHO and OGLE lightcurves
exist for many counterparts (e.g. Coe \etal\ 2008,
McGowan \etal\ 2008b).
The overall X-ray
pulsar properties can tell us about the evolutionary similarities and
differences of a very nearby galaxy compared to our own.

\end{document}